\documentstyle [12pt,epsf]{article}

\newcommand{\beq}{\begin{equation}}

\newcommand{\eeq}{\end{equation}}
\newcommand{\bea}{\begin{eqnarray}}
\newcommand{\eea}{\end{eqnarray}}

\newcommand{\vf}{\varphi}
\newcommand{\Q}{\tilde{Q}_{_L}}
\newcommand{\q}{\tilde{q}_{_R}}
\newcommand{\Lp}{\tilde{L}_{_L}}
\newcommand{\lp}{\tilde{l}_{_R}}

\newcommand{\nL}{$\, \backslash \! \! \! \! L \ $}

\begin{document}
\baselineskip 7.3 mm

\def\thefootnote{\fnsymbol{footnote}}

\begin{flushright}
\begin{tabular}{l}
CERN-TH/97-106\\
hep-ph/9705361
\end{tabular}
\end{flushright}

\vspace{8mm} 

\begin{center}

{\Large \bf 
Phase transitions precipitated by solitosynthesis 
}
\\ 
\vspace{8mm}

\setcounter{footnote}{0}

Alexander Kusenko\footnote{ email address:
kusenko@mail.cern.ch} \\
Theory Division, CERN, CH-1211 Geneva 23, Switzerland \\

\vspace{12mm}

{\bf Abstract}
\end{center}

Solitosynthesis of Q-balls in the false vacuum can result in a phase
transition of a new kind.  Formation and subsequent growth of Q-balls via
the charge accretion proceeds until the solitons reach a critical charge,
at which point it becomes energetically favorable for the Q-ball interior
to expand filling space with the true vacuum phase.   Solitosynthesis can
destabilize a false vacuum even when the tunneling rate is negligible.  
In models with low-energy supersymmetry, where the Q-balls  associated with
baryon and lepton number conservation are generically present,
solitosynthesis can precipitate transitions between the  vacua with
different VEV's of squarks and sleptons.  

\vspace{30mm}

\begin{flushleft}
\begin{tabular}{l}
CERN-TH/97-106 \\
May, 1997 
\end{tabular}
\end{flushleft}

\vfill

\pagestyle{empty}

\pagebreak

\pagestyle{plain}
\pagenumbering{arabic}
\renewcommand{\thefootnote}{\arabic{footnote}}
\setcounter{footnote}{0}

\pagestyle{plain}

A variety of models, in particular the supersymmetric generalizations of
the Standard Model (SSM), possess non-topological solitons of the Q-ball
type in their spectrum \cite{ak_mssm}.  Their existence and stability 
is due to some conserved charge associated with a global U(1) symmetry,
{\it e.\,g.}, baryon (B) or lepton (L) number.  In the early 
Universe, Q-balls can be created \cite{s_gen} in the course of a phase
transition  (``solitogenesis''), or they can be produced via fusion 
\cite{foga,gk} in a process reminiscent of the big bang nucleosynthesis
(``solitosynthesis'').   Finally, small Q-beads \cite{ak_qb} 
can be pair-produced at high temperature. 

In the false vacuum, Q-balls can have a negative energy density in their
interior if there is a lower U(1)-breaking minimum in the potential, 
or, more generally, whenever the scalar potential is negative for some 
value of a charged field. 
When the charge of such Q-ball reaches a critical value, it becomes
energetically favorable for the soliton interior to expand filling the
Universe with the ``true vacuum'' phase.   A first-order phase transition
facilitated by a critical-size Q-ball differs from tunneling in that 
(i) the ``critical bubble'' of the true vacuum is charged and has different
symmetries from those of the usual ``bounce''; and, most importantly, 
(ii) the critical Q-ball can build up gradually via the charge accretion.  
The role of  charge conservation is to stabilize the sub-critical Q-balls
and prevent them from collapsing while they grow in size gradually until
their charge reaches the critical value $Q_c$.  
This allows for a new type of a phase transition, that precipitated
by solitosynthesis.  

This type of a phase transition is different from that previously 
discussed in the literature \cite{coleman1,spector,ellis}, where the
Q-balls form at high temperature,  before the vacuum becomes metastable,
and grow in size slowly as the temperature decreases because of the changes
in the temperature-dependent effective potential.  In this letter we
consider solitosynthesis that takes place at relatively low temperatures in
the false vacuum whose lifetime can be large on the cosmological time scale.  
When a Q-ball reaches a critical size $Q_c \stackrel{>}{_{\scriptstyle
\sim}} 10^2$, it can facilitate a first-order phase transition which would
otherwise never take place during the lifetime of the Universe because the 
probability of tunneling is too small. 

Copious production of critical Q-balls is possible if a chain of reactions
that leads from individual particles to the critical-size solitons remains
in thermal equilibrium.  A freeze-out of any one of these processes can 
stymie the solitosynthesis.  It is, therefore, important that, unlike other
non-topological solitons, Q-balls have no minimal charge \cite{ak_qb}. The
existence of small Q-balls allows for an equilibrium production of the 
large-charge solitons through the charge accretion onto the small ones. 

As an example, we point out that in the MSSM, where the B- and L-balls
are generically present \cite{ak_mssm}, solitosynthesis can precipitate an
otherwise impossible phase transition from an electric charge breaking
to the standard vacuum.  In determining constraints (see, {\it
e.\,g.}, Ref. \cite{kls} and references therein) 
on the MSSM parameters from color and charge breaking minima, one has to
take into account this possibility.

\section{Q-balls in the false vacuum}

Let us consider a complex scalar field $\vf$ with a potential $U(\vf)$ 
that has a local minimum at $\vf=0$ and is invariant under a U(1) symmetry
$\vf \rightarrow \exp(i\theta) \vf$.  We also require that $U(\vf)$ admit
Q-balls, that is \cite{coleman1} 

\beq
U(\vf) \left/ |\vf|^2 \right. = {\rm min},
\ \ {\rm for} \ 
\vf=\vf_0 \neq 0. 
\label{condmin}
\eeq
Then a Q-ball solution \cite{coleman1} of the form
$\vf(x,t)=e^{i\omega t} \bar{\vf}(x)$, where $\bar{\vf(x)}$ is real and
independent of time, minimizes energy for a fixed charge

\beq
Q= \frac{1}{2i} \int \vf(x,t)^* \stackrel{\leftrightarrow}{\partial}_t  
\vf(x,t) \, d^3x \ = \omega \int \bar{\vf}^2(x) \ d^3x.
\label{Q}
\eeq

The energy (mass) of the Q-ball can be written as 

\beq
E(Q) = \frac{Q^2}{ 2 \int  \bar{\vf}^2 d^3x} \ + \ 
T \ + \  V, 
\label{E} 
\eeq
where $T= \int \frac{1}{2} (\nabla \bar{\vf})^2 \ d^3x $
is the gradient energy, and $V=\int U(\bar{\vf}) d^3x $ is the potential
energy of the Q-ball. 

A large Q-ball is a spherical object, inside which the field is close to 
$\vf=\vf_0$ defined in equation (\ref{condmin}), while $\vf=0$ outside. 
We now assume that $U(\vf_0)<0$, so that $V<0$ for a large enough Q-ball. 

Let us examine the energy of a fixed-charge field configuration as a
function of its size using a one-parameter family of the test functions 
obtained from the Q-ball solution $\bar{\vf}$ by expanding (contracting) it
by a factor $\alpha$:  
 
\beq
\vf_\alpha(x) = \bar{\vf}(\alpha x)
\label{alpha}
\eeq
The corresponding energy is 

\beq
E_\alpha (Q) = \frac{1}{\alpha^3} \, 
\frac{Q^2}{ 2 \int  \bar{\vf}^2 d^3x} \ + \ 
\alpha \, T +  \alpha^3 \, V, 
\label{Ealpha} 
\eeq
where $T>0$, but $V<0$.  By definition, $E_\alpha$ has a minimum at
$\alpha=1$.  Therefore, $(d E_\alpha/d\alpha)|_{\alpha=1} =0$.  However, 
in general the equation $(d E_\alpha/d\alpha)=0$ can have two roots, the 
one corresponding to a minimum ($\alpha_1=1$), and the 
one that corresponds to a maximum ($\alpha_2 > 1$).  

The $\alpha=1$ solution, $\bar{\vf}(x)$, is the Q-ball and is a 
(local) minimum of energy.  The $\alpha=\alpha_2$ solution is unstable. 
It represents a kind of a Q-bounce, a critical bubble of the true vacuum, 
such that $\vf_\alpha(x)$ will expand indefinitely for $\alpha>\alpha_2$,
or will contract to a Q-ball for $\alpha<\alpha_2$.

\begin{figure}
\setlength{\epsfxsize}{3.3in}
\centerline{\epsfbox{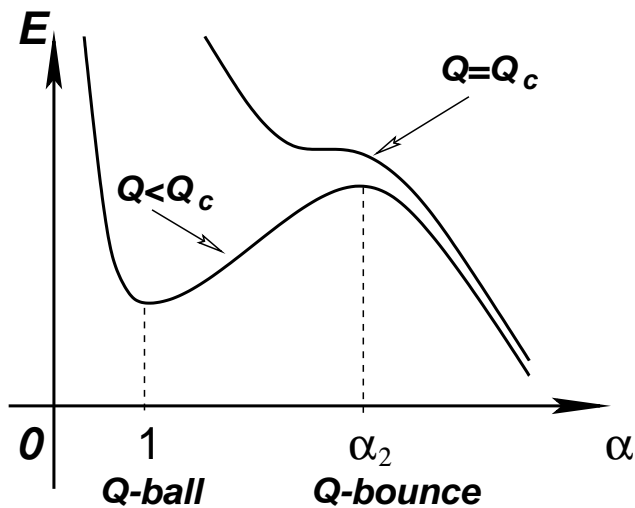}}
\caption{
The energy of a fixed-charge solution  in the false vacuum as a function of
its size.   
}
\label{fig1}
\end{figure}

If $Q$ is small, $E_\alpha(Q)$ in equation (\ref{Ealpha}) has both a
minimum (Q-ball) and a maximum (Q-bounce) as shown in Fig. 1.   
However, as the charge $Q$ approaches 
a critical value $Q_c$, the minimum and the maximum come closer together 
and coincide for $Q=Q_c$.  At this point, there is only one non-trivial
field configuration.  It is unstable.  
After a critical $Q_c$-ball is formed, it will expand filling the space
with the true vacuum, much like a critical bubble in the event of
tunneling.  The value of the critical charge is given by 

\beq
\left. \frac{d E_\alpha}{d\alpha} \right |_{\alpha=1} = 
\left. \frac{d^2 E_\alpha}{d\alpha^2} \right |_{\alpha=1} = 0,
\label{dE2da}
\eeq
and is defined implicitly by the relation 

\beq
Q_c=
\left [ -\int U[\bar{\vf}(x)] d^3x 
\right ]
\left [ \int \bar{\vf}^2(x) d^3x 
\right ].
\label{Qc_imp}
\eeq
Equation (\ref{Qc_imp}) is exact but implicit since the solution
$\bar{\vf}(x)$ is itself a function of $Q_c$.  Through a virial theorem
\cite{ak_qb}, it also implies that, for a critical charge Q-ball, the
identity holds: $2T=-9V$. 

To obtain an explicit formula for the critical charge $Q_c$, we will use
the thin-wall approximation \cite{coleman1}, which is valid in the limit of
large $Q$. (In what follows we will consider cases with $Q_c \gg 1$ only.)  
In the thin-wall limit, 
a Q-ball is approximated by a spherical bubble of radius $R$ with 
$\vf(r) =\vf_0$ inside, for $r \equiv \sqrt{x_i x^i} <R$,  and 
$\vf(r) =0$ for $r>R$.  The thin-wall approximation is good if 
$|U(\vf_0)| \ll (S_1)^{4/3}$, where $S_1 = \int_0^{\vf_0}\sqrt{2 
\hat{U}_\omega(\vf)} d \vf$, $\hat{U}_\omega(\vf)= U(\vf)-(\omega^2/2) 
\vf^2$, and $\omega = Q/(\frac{4}{3} \pi R^3 \vf_0^2)$.  A large charge
requires a small value of $\omega$ \cite{ak_qb}, so that $U(\vf)\approx
\hat{U}_\omega (\vf)$ for large $Q$ in the false vacuum\footnote{
This is not the case in the true vacuum, where $\omega \rightarrow
\omega_0>0$ for $Q \rightarrow \infty$ \cite{coleman1,ak_qb}.  In the false
vacuum $\omega_0=0$.  Therefore, for large $Q$ one can approximate $S_1$ by
its value at $\omega=0$.}. 

In the  thin-wall limit, the Q-ball energy is 

\beq
E(Q) = - |U(\vf_0)| \left (\frac{4}{3} \pi R^3 \right ) + 4\pi R^2 S_1 +
\frac{Q^2}{2 \vf_0^2 \, (4/3) \pi R^3}.
\label{ER}
\eeq
The charge $Q=Q_c$ when $dE/dR = 0$ and $d^2E/dR^2 = 0$ simultaneously. 
From equation (\ref{ER}), 

\beq
Q_c= \frac{100 \pi \sqrt{10}}{81} \, \frac{\vf_0 S_1^3}{U_0^{5/2}},
\label{Qc}
\eeq
where $U_0 \equiv -U(\vf_0) > 0$. 

One can ask how large should the critical charge $Q_c$ be to destabilize
a false vacuum which is otherwise stable on the cosmological time
scale.  We will consider two cases: (a)~a metastable false vacuum at zero
temperature with a lifetime that exceeds the age of the Universe, and 
(b)~a finite-temperature case: a false vacuum at $T>0$, whose decay
width is small in comparison to the rate of expansion of the Universe. 
At zero temperature, the decay probability of the false vacuum per unit 
four-volume is 

\beq
\Gamma/{\sf V} = A \, e^{-B}, 
\label{gamma}
\eeq
where $B$ is the Euclidean action of the bounce \cite{tunn}. 

Suppose the Universe rests in the false vacuum whose lifetime exceeds 
$t_0\sim 10^{10}$ years.  This implies that $\Gamma/{\sf V} \ll 1$ for 
${\sf V} \sim t_0^4$. Taking the pre-exponential factor to be of order
$A\sim (100 \, {\rm GeV})^4$, one obtains the constraint 
$B \stackrel{>}{_{\scriptstyle \sim}} 400$ for a false vacuum to 
be considered stable. 

On the other hand, in the thin-wall approximation, the action of the bounce
associated with tunneling into a global minimum at $\vf=\vf_0$ is
\cite{tunn} 

\beq
B= \frac{27 \pi^2}{2} \, \frac{S_1^4}{U_0^{3}}. 
\label{B}
\eeq
From equations (\ref{Qc}) and (\ref{B}), 

\beq
Q_c= \frac{200}{729} \sqrt{\frac{5}{\pi}} \, \left ( \frac{2 B^3}{3} 
\right )^{1/4} 
\frac{\vf_0 }{U_0^{1/4}}
\label{QcB}
\eeq

For $B\sim 400$, $Q_c \sim 28 (\vf_0/U_0^{1/4})$ is sufficient to
destabilize the vacuum.  Similarly, for the first-order phase transition at
finite temperature \cite{linde},  

\beq
\Gamma/{\sf V} = A(T) \, e^{-S_3/T}, 
\label{gammaT}
\eeq
where $S_3$ is the tree-dimensional bounce action, for which the thin-wall
approximation yields \cite{linde} $S_3=16 \pi S_1^3/(3 U_0^2(T))$.  The
critical charge can once again be re-expressed in terms of the bounce
action and the (now temperature-dependent) depth of the true vacuum:

\beq
Q_c=\frac{25}{54} \sqrt{\frac{5}{2}} \, \frac{\vf_0 S_3}{U_0^{1/2}(T)}
\label{QcT}
\eeq
Tunneling rate is negligible in comparison to the expansion rate of the
Universe at the electroweak scale temperatures if $S_3/T  
\stackrel{>}{_{\scriptstyle \sim}} 200$.  A Q-ball with a charge $Q> Q_c 
\sim 146 \vf_0 T/U_0^{1/2}(T)$ will facilitate an otherwise impossible 
phase transition in such a system. 

In the next section, we discuss the conditions under which Q-balls of
critical size can be produced in thermal equilibrium via the accretion of
charge.  The result is a rapid phase transition that proceeds via
nucleation of charged critical bubbles of the true vacuum.

\section{Synthesis of critical Q-balls}

Solitosynthesis \cite{foga,gk} of large Q-balls in (approximate) thermal
equilibrium is possible as long as the chain of requisite reactions leading
from small to large solitons is not hampered by freeze-out.  
First, we require that the equilibrium number density of the $\vf$
particles is maintained via some microscopic processes,  

\beq
\vf \bar{\vf} \longleftrightarrow {\rm light \ particles}, 
\label{reaction}
\eeq
which are sufficiently fast in comparison to the expansion rate of the
Universe.   

Second, the following chain of reactions is essential for the
solitosynthesis: 

\begin{eqnarray}
\vf + \vf & \longleftrightarrow  & \Phi(2) \nonumber \\ 
{\rm a \ few} \ \vf'{\rm s} & \longleftrightarrow & \Phi({\rm few})
\label{firststep} \\
\nonumber \\
& \cdot \ \cdot \ \cdot & \nonumber \\
\Phi(Q) + \vf & \longleftrightarrow & \Phi(Q+1) \label{nextstep} \\
\nonumber \\
& \cdot \ \cdot \ \cdot , & \nonumber 
\end{eqnarray}
where $\Phi(Q)$ denotes a soliton with the charge $Q$. 

It is crucial that, unlike some other non-topological solitons, 
classically stable Q-balls exist for very small (integer) 
charges $Q \ge 1$ \cite{ak_qb}.  If the minimal charge of a soliton 
$Q_{min} \gg 1$, then the chain of solitosynthesis reactions would have to 
start from a process $\vf+\vf+...+\vf \leftrightarrow \Phi(Q_{min})$ 
that is increasingly rare for large $Q_{min}$ and cannot be in
equilibrium.  Synthesis of non-topological solitons with $Q_{min} \gg 1$ 
is stymied by the lack of the first step reaction that could produce the
seeds for subsequent accretion.  
In the case of Q-balls, there is no minimal charge\footnote{
A naive application of the thin-wall formuli to small Q-balls, 
outside the thin-wall limit,  could lead one to an erroneous conclusion
that Q-balls must have a large charge to be stable.  This apparent 
constraint is merely an artifact of the thin-wall approximation
\cite{ak_qb}.}, except the quantization condition requires that $Q$ be an
integer; hence $Q \ge 1$.   The first of these reactions (\ref{firststep}) 
may also be supplemented by the pair production of small solitons
via charge fluctuations \cite{gkm}. A freeze-out temperature of the
reactions (\ref{reaction}) --  (\ref{nextstep}), $T_f$, is an important 
model-dependent parameter. 

In chemical equilibrium,  reactions (\ref{firststep}) -- (\ref{nextstep})
enforce a relation between the chemical potentials of $\vf $ 
and $\Phi (Q)$: $\mu (Q) = Q \mu(\vf)$.  This allows one to express the
$Q$-ball number density $n_{_Q}$ in terms of the $\vf$-number density, 
$n_\vf $ (see, {\it e.\,g.}, Ref. \cite{foga}): 

\beq
n_{_Q} = \frac{g_{_Q}}{g_\vf^Q} n_\vf^Q \, \left ( \frac{E(Q)}{m_\vf} 
\right )^{3/2} \,  \left ( \frac{2\pi}{m_\vf T} \right )^{3(Q-1)/2} \, 
\exp \left ( \frac{B_{_Q}}{T} \right ),
\label{nQ}
\eeq
where $B_{_Q}= Q m_\vf - E(Q)$, $g_{_Q}$ is the internal partition
function of the soliton, and $g_\vf$ is the number of degrees of freedom
associated with the $\vf$ field.  A  soliton has $B_{_Q}>0$.  Charge
conservation implies:  

\beq
n_\vf = \eta_\vf n_\gamma - \sum_Q Q n_{_Q}, 
\label{Qconst}
\eeq
where $\eta_\vf $ is the Q-charge asymmetry.  Here we have assumed that
there is no light fermions or vector bosons carrying the same $U_{_Q}(1)$
charge.  In the presence of light charged particles, the charge asymmetry
$\eta$ can be accommodated by the light particle densities, allowing
$n_\vf$ to drop to its $\mu \ll m_\vf$ equilibrium value suppressed by
$\exp (-m_\vf/T)$,  which is too low for the solitosynthesis to take
place\footnote{The author thanks M. Shaposhnikov for illuminating 
discussions of this and other issues related to solitosynthesis.}.

Equations (\ref{nQ}) and (\ref{Qconst}), usually solved numerically, 
describe an explosive growth of solitons below some critical temperature
$T_s$ \cite{foga,gk}.  This exponentially fast growth is hampered 
eventually, when the population of the $\vf$ particles is depleted in
accordance with the charge conservation (\ref{Qconst}). One can show that
the growing population of  Q-balls is dominated by those with a large
charge.   

If a single soliton with a critical charge $Q_c$ is produced per Hubble
volume, a phase transition takes place.  In fact, at $T=T_s$
critical solitons are copiously produced.  We will obtain a crude
analytical estimate of $T_s$, the temperature at which $n_{Q_c}$ starts 
growing.  

Based on the results of the preceding section, we take $Q_c \gg 1 $.  
In the limit of large $Q$, $g_{_{Q+1}}/g_{_Q} \approx 1$, $E(Q+1)/E(Q) 
\approx 1$, and one can write the Saha equation in the form 

\beq
\frac{d}{dQ} (\ln n_{_Q}) = \frac{n_{_{Q+1}}}{n_{_Q}} -1 = 
\left ( \frac{n_\vf}{g_\vf}  \right )
\left ( \frac{2\pi}{m_\vf T} \right )^{3/2} \exp
\left ( \frac{b_{_Q}}{T} \right) -1, 
\label{saha_diff}
\eeq
where $b_{_Q} = m_\vf -[E(Q+1)-E(Q)] \approx  m_\vf - dE(Q)/dQ$. 

Differential equation (\ref{saha_diff}) describes a growing (with $Q$) 
equilibrium population of large solitons when its right-hand side is
positive.  We wish to determine 
the temperature $T_s$, at which the growth of critical Q-balls begins. 
For the initial growth of $n_{_{Q_c}}$, we neglect the back reaction 
on the $\vf$ population and take $n_\vf = \eta_\vf n_\gamma \sim 
\eta_\vf T^3$ in accordance with equation (\ref{Qconst}).  The right-hand
side of equation (\ref{saha_diff}) is positive for temperatures below $T_s$
defined by  

\beq
T_s = \frac{b_{_Q}}{|\ln \eta_\vf |+(3/2) \ln (m_\vf/T_s)-\ln g_\vf}. 
\label{Ts1}
\eeq

A copious production of critical solitons is possible if  

\beq
T_s > T_f.
\label{cond}
\eeq
As was explained earlier, another condition for solitosynthesis is that all
the particles that carry the $U_{_Q}(1)$ charge, including fermions and gauge
bosons,  must have masses that are large in comparison to $T_s$.
For scalars this condition is satisfied automatically. 

\section{Phase transitions in supersymmetric models}

A particularly interesting example is 
baryon (B) and lepton (L) balls in the MSSM \cite{ak_mssm}. We will
continue to use $\vf$ as a generic name for the squarks and sleptons.  
A $B$- or $L$- ball can grow  via absorption of the SU(2)-doublet sleptons
$\Lp$ and squarks  $\Q$, as well as the corresponding singlets $\q$ and
$\lp$. Their equilibrium population is maintained through several
reactions,
whose total cross-section $\sigma$, and hence the freeze-out temperature 
$T_f \approx m_\vf/\ln (m_\vf m_{_P}\sigma)$ depend on the parameters of
the model.  

A typical cross-section of the processes (\ref{nextstep}) is
characterized by a relatively large Q-ball size, $\sigma \sim 4\pi
R_{_Q}^2 $.  These reactions, therefore, are not expected to 
freeze out before the annihilation reactions (\ref{reaction}).  Therefore,
the relevant freeze-out temperature is determined by the total annihilation
cross-section $\sigma$ of the squarks (sleptons). 

As an example, let us consider a local lepton number breaking (\nL) minimum 
that develops along the $H_1 \Lp \lp \neq 0 $ direction \cite{clm} due to
the corresponding (supersymmetric)  tri-linear coupling 
proportional to the $\mu$-term\footnote{Although in the preceding section 
$\mu$ was used to denote a chemical potential, we trust that no confusion
will arise as we adopt the common notation for the $\mu$-term in the MSSM.}.
It arises from the superpotential  $W=\lambda H_2 L_{_L} l_{_R} + \mu H_1
H_2+...$  The analysis of 
the MSSM scalar potential shows that both $H_1$ and $H_2$ have large VEV's
in this minimum, $\langle H_{1,2} \rangle \sim \mu/\lambda = \tilde{v}$,
where $\lambda \mu $ is the corresponding tri-linear coupling and $\lambda
$ is the associated Yukawa coupling.  Both $\Q$ and $\q$ have vanishing VEV
in the \nL minimum.  In the standard vacuum, $H_1^2 + H_2^2 =
v^2$, where $v \approx 250$ GeV.  

\begin{figure}
\setlength{\epsfxsize}{3.3in}
\centerline{\epsfbox{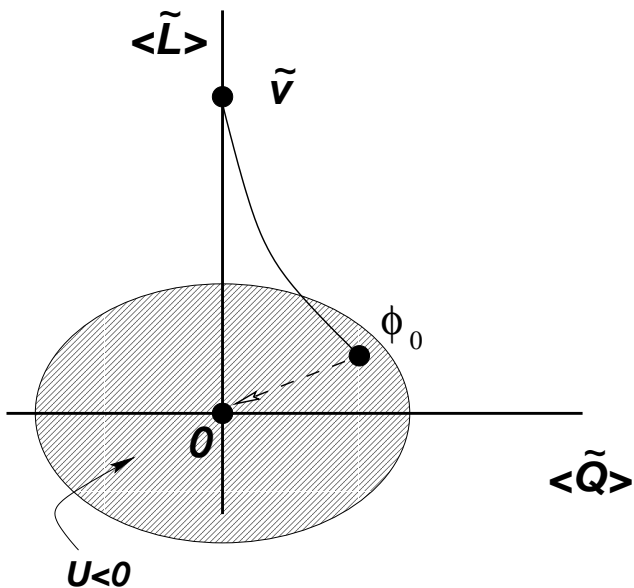}}
\caption{
The energy of a large baryonic ball is minimized when {$\vf(x)=\vf_0$}, 
{$U(\vf_0)<0$}, in its interior.  Although both the true ({$\Lp=0$}) and the 
false ({$\Lp=\tilde{v}$}) vacua conserve the baryon number, {$B$}-balls
precipitate a phase transition into a negative-energy attractive domain
of the true vacuum (shaded region).  
}
\label{fig2}
\end{figure}

We consider the case of a metastable \nL vacuum decaying into a deeper 
standard (lepton number conserving) true vacuum (Fig. \ref{fig2}).  Both
vacua conserve the baryon number since $\langle \Q  \rangle =\langle \q
\rangle = 0$, and because the $SU(2)\times U(1)$ gauge group is broken, so  
the sphaleron transitions are suppressed. For simplicity, we will assume
that there is no other false vacuum with energy density below that of the
\nL minimum. 

If the Yukawa coupling $\lambda$ is small and $\mu$ is of the order the 
electroweak scale, then $\tilde{v}/v \gg 1$ and the \nL vacuum is
``very far'' from the true vacuum.  The tunneling rate is then
suppressed by the huge size of the barrier and is negligible on the
cosmological time scale (the probability of tunneling goes, roughly, as
$\exp(-const/\lambda^2)$ \cite{ehnt}).  

We will show that for $\tilde{v}/v \gg 1$, a rapid transition from the 
false \nL vacuum to the standard true vacuum can be precipitated by  
the solitosynthesis of baryonic balls.  

It is easy to see that there is always a value of charge (baryon number, in
our example) $Q_1$, such that for $Q>Q_1$ the energy (\ref{E}) of a  B-ball
in the false vacuum is minimal when the potential energy in its 
interior is negative ($V<0$).  Since the origin (Fig. \ref{fig2}) is a
global minimum, there is a domain with a negative energy density in the
field space around the origin.  This domain is shown as a shaded region in 
Fig. 2.  By assumption, $U(\vf) \ge 0$ everywhere
outside this region.  A nucleation of a critical B-ball with $\vf(0)=\vf_0,
\ U(\vf_0)<0$, will, therefore,
convert the false vacuum  into a phase that will eventually relax to the 
true vacuum at the origin.  From this example one can see, in particular, 
that Q-balls associated with some $U_{_Q}(1)$ symmetry can mediate phase 
transitions between the two vacua, both of which preserve $U_{_Q}(1)$. 

As follows from the discussion in the last section, solitosynthesis of
baryonic balls can take place only in the absence of light baryons. 
For a large enough Higgs VEV in the false vacuum, the the masses of  
quarks will be approximately equal to those of squarks.  The mass
splittings are due to the soft supersymmetry breaking terms 
independent of the Higgs VEV.  Since the Higgs VEV in the \nL vacuum
associated with  a small Yukawa coupling $\lambda$ is large ($\sim
\mu/\lambda$),  it can be much greater than the SUSY breaking terms
responsible for the mass difference.  As a result, the quark masses are of
order the squark masses in the false vacuum, where the typical scale $m_\vf
\sim \mu/\lambda$ can be as large as $10^3...10^6$ GeV. 

We will examine condition (\ref{cond}) for $m_\vf \sim 1$ TeV in the false 
\nL vacuum (while $m_\vf \sim 10^2$ GeV in the true vacuum). 
We take the baryon (lepton) asymmetry $\eta_\vf \sim 10^{-10}$.  
Production of critical solitons can take place at the temperature 

\beq
T_s 
\approx b_{_Q}/\ln(1/\eta_\vf) \approx \frac{1}{23} 
\left ( m_\vf - \left. \frac{d}{dQ} E(Q) \right |_{Q=Q_c} \right )
\eeq
if 

\beq
T_s > \  T_f \approx m_\vf/\ln(m_\vf m_{_P} \sigma) \approx m_\vf/40
\label{cond_mssm}
\eeq
One can easily check that the charge accretion is many orders of
magnitude faster than the Hubble time scale at these temperatures. 
The characteristic time scale associated with the growth of a Q-ball 
with the charge $Q\sim Q_c$ can be estimated as 

\beq
\tau_{requir}^{-1}  \sim \frac{dQ}{dt} \sim n_\vf \sigma v_\vf \sim 
\eta_\vf n_\gamma (4 \pi R_{Q_c}^2) \sqrt{T_s/m_\vf} \sim 
10^{-7} GeV
\label{time1}
\eeq
while the Hubble scale at these temperatures is $\tau_{avail}^{-1} 
\sim T_s^2/m_{_P} \sim 10^{-15}$ GeV.   Clearly, $\tau_{requir} \ll 
\tau_{avail}$, hence the Q-balls have plenty of time to grow when the 
condition (\ref{cond_mssm}) is satisfied.  

We note in passing that although $dE/dQ$ is a constant for large Q-balls 
in the true vacuum \cite{coleman1}, this is generally 
not the case for Q-balls in the false vacuum.  For instance, in the limit
$U_0 \rightarrow 0$, $E(Q) \propto  Q^{4/5}$ \cite{spector}. 

In summary, the first-order phase transition of a new kind can result 
from the solitosynthesis of Q-balls (in particular, B- and L-balls in the
MSSM), which can destabilize a metastable vacuum even when the tunneling
rate is negligible.  At a certain temperature 
(\ref{Ts1}),  Q-balls build up rapidly by absorbing charged particles
from the outside until the critical size is reached, at which point they
expand and fill the Universe with the ``true vacuum'' phase.  This
mechanism has important implications for models with low-energy
supersymmetry, in which the presence of baryonic and leptonic Q-balls is
generic.   

The author thanks M. Shaposhnikov for many stimulating discussions.

\end{document}